\documentclass[12pt,chap,a4paper]{article}
\usepackage{jheppub}
\pdfoutput=1
\usepackage{graphicx}
\usepackage{amssymb,latexsym} %Fonts
\usepackage{graphics,subfigure,graphicx}
\usepackage{amsmath,xspace}
\usepackage[compat=1.1.0]{tikz-feynman}
     
\usepackage{eepic}
\newcommand{\Herwig}{\textsf{Herwig}\xspace}

\usepackage{array}
\newcolumntype{L}[1]{>{\raggedright\let\newline\\\arraybackslash\hspace{0pt}}m{#1}}
\newcolumntype{C}[1]{>{\centering\let\newline\\\arraybackslash\hspace{0pt}}m{#1}}
\newcolumntype{R}[1]{>{\raggedleft\let\newline\\\arraybackslash\hspace{0pt}}m{#1}}

\newcommand{\opm}{ 
  \mathbin{
    \mathchoice
      {\buildcirclepm{\displaystyle     }{0.14ex}{0.95}{0.05ex}{.7}}
      {\buildcirclepm{\textstyle        }{0.14ex}{0.95}{0.05ex}{.7}}
      {\buildcirclepm{\scriptstyle      }{0.13ex}{0.955}{0.04ex}{.55}}
      {\buildcirclepm{\scriptscriptstyle}{0.08ex}{0.95}{0.03ex}{.45}}
  } 
}

\newcommand\buildcirclepm[5]{%
  \begin{tikzpicture}[baseline=(X.base), inner sep=-#5, outer sep=-.65]
    \node[draw,circle,line width=#4] (X)  {\footnotesize\raisebox{#2}{\scalebox{#3}{$#1\pm$}}};
  \end{tikzpicture}%
}

\newcommand{\omp}{ 
  \mathbin{
    \mathchoice
      {\buildcirclemp{\displaystyle     }{0.14ex}{0.95}{0.05ex}{.7}}
      {\buildcirclemp{\textstyle        }{0.14ex}{0.95}{0.05ex}{.7}}
      {\buildcirclemp{\scriptstyle      }{0.13ex}{0.955}{0.04ex}{.55}}
      {\buildcirclemp{\scriptscriptstyle}{0.08ex}{0.95}{0.03ex}{.45}}
  } 
}

\newcommand\buildcirclemp[5]{%
  \begin{tikzpicture}[baseline=(X.base), inner sep=-#5, outer sep=-.65]
    \node[draw,circle,line width=#4] (X)  {\footnotesize\raisebox{#2}{\scalebox{#3}{$#1\mp$}}};
  \end{tikzpicture}%
}

\title{
Initial State Radiation in the \Herwig{}~7 Angular-Ordered Parton Shower
}
\author[a]{Gavin~Bewick}
\author[a,b]{Silvia~Ferrario~Ravasio}
\author[a,c]{Peter~Richardson}
\author[d]{Michael~H.~Seymour}
\affiliation[a]{Institute for Particle Physics Phenomenology, Durham
  University}
\affiliation[b]{Rudolf Peierls Centre for Theoretical Physics, University of Oxford, Oxford, UK}
\affiliation[c]{Theoretical Physics Department, CERN, Switzerland}
\affiliation[d]{Lancaster-Manchester-Sheffield Consortium for Fundamental Physics, Department of Physics and Astronomy, University of Manchester, M13 9PL, U.K.}
\emailAdd{gavin.bewick@durham.ac.uk}
\emailAdd{silvia.ferrarioravasio@physics.ox.ac.uk}
\emailAdd{peter.richardson@durham.ac.uk}
\emailAdd{Michael.Seymour@manchester.ac.uk}

\abstract{We study the simulation of initial-state radiation in
  angular-ordered parton showers in order to investigate how different
  interpretations of the ordering variable affect the logarithmic
  accuracy of such showers. This also enables us to implement a recoil
  scheme which is consistent between final-state and initial-state
  radiation.  We present optimal values of the strong coupling and
  intrinsic transverse momentum to be used in each version of the
  parton shower, tuned using $Z^0$-boson production at the LHC at 7
  TeV.  With these tuned showers, we perform a phenomenological study
  of the Drell-Yan process at several centre-of-mass energies.}

\preprint{\begin{flushright}CERN-TH-2021-103\\OUTP-21-18P\\MCnet-21-13\\IPPP/21/05\end{flushright}}

\begin{document}

\maketitle
\flushbottom

\section{Introduction}

The complex hadronic final states of high energy particle collisions require the use of General Purpose Monte Carlo (GPMC) event generators to turn the fundamental theory into predictions of the observable final states. These event generators take a calculation of a hard scattering process and evolve it down in scale using a parton shower (PS) algorithm to a low energy scale where perturbation theory breaks down and a hadronization model is used to convert the coloured partons into hadrons, which can then also decay using a non-perturbative model. A model of Multiple Parton Interactions (MPI) is also required to describe hadron-hadron collisions.

Research in this area over the last decade has focused on matching the parton shower with fixed-order hard processes, but there have also been recent attempts to improve the parton shower algorithms themselves. It was one such work \cite{Dasgupta:2018nvj} on the problems dipole
showers\footnote{This work focused on the {\sf PYTHIA 8}~\cite{Sjostrand:2004ef} and {\sf Dire}~\cite{Hoche:2015sya} dipole showers but the issues raised apply to all dipole showers.} face even at leading-logarithmic accuracy that prompted us to make similar analyses of the \Herwig{}~{\sf 7} angular-ordered parton shower~\cite{Bewick:2019rbu}. 

Logarithmic accuracy can be defined in terms of how well the PS reproduces the singular limits of soft-collinear matrix elements of QCD. In the limit where gluon emission is widely separated in angle and energy then for $n$ emissions there will be $2n$ large logarithms, $L$, and the emission probability is proportional to $\alpha_s^n L^{2n}$. These are the Leading Logarithmic (LL) contributions. For emissions which are collinear \textbf{or} occur at commensurate energies the emission probability is proportional to  $\alpha_s^n L^{2n-1}$ and these are the next-to-leading logarithmic contributions. Parton shower algorithms are formally only LL accurate, but by using quasi-collinear splitting functions, the transverse momentum as the scale of the strong coupling and the Catani-Marchesini-Webber~(CMW)~\cite{Catani:1990rr} scheme for the two-loop coupling, they are also able to account for LL~(leading logarithmic) and NLL~(next-to-leading logarithmic) contributions with the exception of soft wide-angle gluon emission.

In Ref.~\cite{Bewick:2019rbu} we examined how the choice of recoil scheme in
the shower affects the logarithmic accuracy for final-state radiation
(FSR). We found that of the three schemes presented, the ``$p_\perp$
preserving'' and the ``dot-product preserving'' schemes both satisfied
conditions to be NLL accurate while the ``$q^2$ preserving'' scheme, which
was the default of \Herwig{}~{\sf 7} prior to that publication, was not NLL
accurate.

In addition to FSR, there is also initial state radiation (ISR) to
consider.  The modelling of ISR can have a large impact on the
transverse momentum distributions of colour-singlet systems produced
in hadron collisions.  The Drell-Yan (DY) process, in which a pair of
leptons is produced in a hadron collision, is an example of a
``standard-candle process'', as the lepton momenta can be measured
precisely to validate our understanding of QCD. Thus, it is of
paramount importance to provide accurate theoretical predictions for
this process.  Issues arising from transverse momentum recoil due to
ISR in dipole showers were addressed in Ref.~\cite{Platzer:2009jq}.
These were mainly related to the fact that a final state singlet, such
as in Drell-Yan production, receives a non-vanishing transverse
momentum contribution only from the very first shower emission.  The
authors proposed to allow the incoming partons to take the transverse
momentum recoil, and then realign them with the beam directions at the
end of the showering phase. Although this solution mitigates the
problem, it is not sufficient to reach NLL accuracy, because the
shower will still face the same recoil-scheme related problems present
for final-state radiation~\cite{Dasgupta:2018nvj}.

In angular-ordered showers these issues are absent, both for the
  initial- and final-state showers, but spurious NLL terms may arise
depending on the interpretation of the ordering variable in the
presence of multiple emissions, as already found in
Ref.~\cite{Bewick:2019rbu} in the context of FSR.  For this reason, in
this work we present the findings of our study into the effects of
recoil scheme on ISR in the angular-ordered shower, focusing on the
schemes which are NLL accurate for FSR. This also enables us to
formulate a recoil scheme which is consistent between FSR and ISR, and
perform phenomenological studies of Drell-Yan processes. In this work
we considered several prescriptions to improve the description of the
hardest emission, and we provided dedicated tunes for all the
considered options.

In Sec.~\ref{sec:kinematics} we describe the implementation of the kinematic mapping for multiple emissions for different interpretations of the ordering variable.
In Sec.~\ref{sec:double-soft} we discuss the radiation pattern for the case of two soft emissions well separated in rapidity. 
The global recoil necessary to ensure full momentum conservation for the production of a colour singlet in hadron-hadron collisions is discussed in Sec.~\ref{sec:global-rec-DY} (more details and the implementation for more generic processes is discussed in appendix~\ref{sec:global_recoil}). Comparisons with experimental data for $Z$- and $W$-boson production are presented in Sec.~\ref{sec:generators}. We present our conclusions in Sec.~\ref{sec:concl}. 

\section{Kinematics}
\label{sec:kinematics}

We consider the emission of initial-state radiation from a parton coming from a proton. Conventionally, we consider such a parton as shower progenitor and evolve the shower backwards from
the scale of the hard emission to the cutoff.
We adopt the Sudakov decomposition for particles such that
\begin{equation}
  p_l = \alpha_l P + \beta_l n + k_{\perp l},
  \label{eqn:sudakov}
\end{equation}
where the reference vectors $P$ and $n$ are the momentum of the parent hadron and a light-like vector which points in the opposite direction.\footnote{In the hadron--hadron frame for colour-singlet processes such as Drell--Yan, or if the colour partner is in the initial state. For processes with an outgoing colour partner, for example DIS, the direction of the colour partner in the Breit frame is used.}
$k_{\perp l}$ is orthogonal to both $P$ and $n$.
For the parton that enters the hard scattering process $\alpha_l = x$, where $x$ is the fraction of the hadron's momentum that enters the hard scattering process.
The parameter $\beta_l$ can be found by imposing that the external particles are on their mass shell.

When we consider an ISR splitting $\widetilde{ij} \to i ,j$ we denote with $i$ the space-like child and with $j$ the time-like one. Space-like partons are always considered massless, while the time-like parton can have a mass $m_j \neq 0$. This means that when a $g \to b \bar{b}$ splitting is considered, one quark will be considered as massless, while the final-state one will be treated as massive.
The parent parton is on-shell and massless and it acquires an increasingly negative virtuality as it emits towards the hard scattering. The momenta obey
  \begin{equation}
P^2=0, \qquad P\cdot n \neq 0, \qquad
n^2=0, \qquad 
P\cdot k_{\perp l} = n\cdot k_{\perp l}=0,
  \end{equation}
so that the transverse momenta are defined relative to the direction of $P$ and $n$. This also means that $k_{\perp l}$ is space-like.

\subsection{One emission}

We consider the case of one emission. If we use the usual Sudakov decomposition the momenta are
\begin{subequations}
  \begin{eqnarray}
    p_{\widetilde{ij}} &=& x P;\\
    p_j &=& (1-z)p_{\widetilde{ij}} + \beta n + p_{\perp}; \\
    p_i &=& p_{\widetilde{ij}} - p_j = z p_{\widetilde{ij}} - \beta n -p_{\perp},
\end{eqnarray}
\end{subequations}
where $z$ is the light-cone momentum fraction
\begin{equation}
z = \frac{p_i \cdot n}{p_{\widetilde{ij}} \cdot n}.
\end{equation}
The coefficient $\beta$ can be determined by requiring that $p_j$ is on its mass-shell
\begin{equation}
\beta = \frac{m^2_j + |p_\perp|^2 }{2(1-z) \,x\, P \cdot n}.
\end{equation}
The virtuality of the space-like parton is
\begin{equation}
p_i^2 = -2 p_j \cdot p_{\widetilde{ij}} + m_j^2 =\frac{-(|p_\perp|^2 + z m_j^2)}{(1-z)}
\label{eq:virt_oneEmission}
\end{equation}
and the ordering variable can be defined as
\begin{equation}
\tilde{q}^2 = \frac{-p_i^2}{1-z} = \frac{2 p_j \cdot p_{\widetilde{ij}} -m_j^2}{1-z}=\frac{(|p_\perp|^2 + z m_j^2)}{(1-z)^2}.
\end{equation}

\subsection{Multiple emissions}

When multiple emissions are considered (\emph{i.e.}~$j$ acquires a positive virtuality $p_j^2$ and/or $\widetilde{ij}$ undergoes a further initial-state splitting), we cannot preserve simultaneously $p_i^2$, $p_j \cdot p_{\widetilde{ij}}$ and $p_{\perp}^2$.

It can be shown that in this case the virtuality of $i$ is
\begin{equation}
p^2_i= -\frac{|p_\perp|^2 + z p_j^2-z(1-z) p^2_{\widetilde{ij}}}{1-z}. 
\label{eq:virt_Generic}
\end{equation}
By comparing Eq.~\eqref{eq:virt_Generic} with Eq.~\eqref{eq:virt_oneEmission}, we notice that the on-shell mass $m_j^2$ has been replaced with the virtuality $p_j^2$ and we have also a term proportional to $p^2_{\widetilde{ij}}$, which is zero when only one emission is concerned.
In order to complete the kinematic reconstruction we thus need to find an expression of $|p_\perp|^2$ as a function of $\tilde{q}^2$, $p_j^2$ and $p_{\widetilde{ij}}^2$.

\subsubsection{$p_\perp$-preserving scheme}
If we preserve the transverse momentum of each emission, we have
\begin{equation}
\tilde{q}^2 \equiv \frac{|p_\perp|^2 + z m_j^2}{(1-z)^2},
\end{equation}
which immediately yields
\begin{equation}
|p_\perp|^2 = (1-z)^2 \tilde{q}^2 - z m_j^2.
\end{equation}

\subsubsection{$q^2$-preserving scheme}
If we preserve the virtuality of the emission, we obtain
\begin{equation}
\tilde{q}^2 \equiv \frac{-p_i^2}{1-z},
\end{equation}
By inverting Eq.~\eqref{eq:virt_Generic} and using the above expression for $p_i^2$ we get
\begin{equation}
|p_\perp|^2 = (1-z)^2\tilde{q}^2 +z(1-z) p_{\widetilde{ij}}^2 - z p_j^2.
\end{equation}
Since $p_{\widetilde{ij}}^2<0$ and $p_j^2>0$, during the evolution $|p_\perp|^2$ will decrease, and can eventually become negative, as already found in the context of FSR~\cite{Bewick:2019rbu}.

\subsubsection{Dot-product preserving scheme}
The final choice we examine is the dot-product preserving scheme:
\begin{equation}
\tilde{q}^2 \equiv \frac{2 p_j \cdot p_{\widetilde{ij}} -m_j^2}{1-z},
\end{equation}
from which we derive
\begin{equation}
2 p_j \cdot p_{\widetilde{ij}} = (1-z) \tilde{q}^2  + m_j^2.
\end{equation}
By inverting Eq.~\eqref{eq:virt_Generic}
and using $p_i^2=(p_{\widetilde{ij}}-p_j)^2$,
\begin{equation}
|p_\perp|^2  =  (1-z)^2 \tilde{q}^2-p_j^2 -(1-z)^2  p_{\widetilde{ij}}^2+(1-z)m_j^2.
\end{equation}
In this case we see that during the evolution of the time-like parton $j$, $|p_\perp|$ is reduced, while during the evolution of the 
space-like parton $\widetilde{ij}$ it is increased. We need to ensure that there is still a physical solution, \emph{i.e.}~$|p_\perp|^2\geq0$,
  following subsequent initial- and final-state radiation. As further ISR can only increase $|p_\perp|^2$ there is no problem, however
  subsequent FSR can reduce $|p_\perp|^2$ and so we must ensure that further final-state radiation satisfies
\begin{equation}
    p_j^2 \leq (1-z)^2 \tilde{q}^2.
\end{equation}
As shown in Ref.\,\cite{Bewick:2019rbu}, the angular-ordering condition for final-state radiation,\linebreak $\tilde{q}_j<(1-z)\tilde{q}$, will
ensure that for further time-like radiation \footnote{See Ref.\,\cite{Bewick:2019rbu} Eqn 4.11 and the surrounding text.}
\begin{equation}
  p_j^2 \leq \frac{\tilde{q}^2_j}{2},
\end{equation}
and therefore the angular-ordering condition ensures that
\begin{equation}
  p_j^2 \leq (1-z)^2 \frac{\tilde{q}^2}{2},
\end{equation}
which ensures there will always be a physical solution for $|p_\perp|^2$.

\subsubsection{Summary}
Remembering that we always consider the case in which $m_i=m_{\widetilde{ij}}=0$, we can summarize all the $|p_\perp|^2$ expressions using
\begin{equation}
|p_\perp|^2 = (1-z)^2(\tilde{q}^2 - P_{\widetilde{ij}}^2) + (1-z)M_j^2 -P_j^2 -(1-z)(m_i^2-M_{\widetilde{ij}}^2),
\label{eq:pT_generic}
\end{equation}
where
\begin{equation}
P^2_l = M^2_l = m^2_l
\end{equation}
in the $p_\perp$-preserving scheme,
\begin{equation}
P^2_l = M^2_l = p^2_l
\end{equation}
in the $q^2$ preserving scheme, and
\begin{equation}
P^2_l = p^2_l, \,\, M^2_l = m^2_l
\end{equation}
in the dot-product preserving scheme.

It can be shown that Eqs.~\eqref{eq:pT_generic} and~\eqref{eq:virt_Generic} are valid also in case of ISR from a resonance. The derivation is identical to the pure ISR case, with the exception that $m_i= m_{\widetilde{ij}} \neq 0$ and that the ordering variable for one emission is
\begin{equation}
\tilde{q}^2 = \frac{-p_i^2 +m_i^2}{1-z} = \frac{2 p_j \cdot p_{\widetilde{ij}} -m_j^2}{1-z}=\frac{|p_\perp|^2 + z m_j^2+(1-z)^2 m_i^2}{(1-z)^2}.
\label{eq:ptDotGeneric}
\end{equation}

\section{Double Gluon Emission}
\label{sec:double-soft}
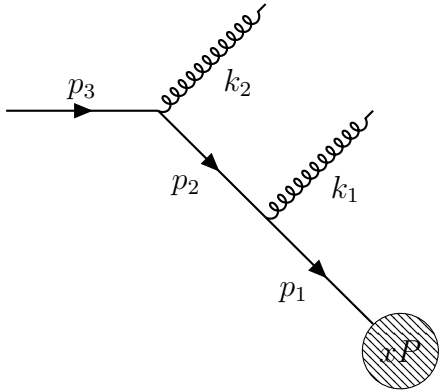
\begin{figure}[h]
\begin{tikzpicture}
  \begin{feynman}[large]
    \vertex (a);
    \vertex [right=of a] (b);
    \vertex [below right=of b] (c);
    \vertex [below right=of c] (d);
    \node[below right=of c, blob] (col) {$xP$};
    \vertex [above right=of b] (g2);
    \vertex [above right=of c] (g1);
 
    \diagram* {
      (a) -- [fermion, edge label = $p_3$] (b) -- [fermion, edge label' = {$p_2$}] (c)-- [fermion, edge label' = {$p_1$}] (d),
      (b) -- [gluon, edge label' = $k_2$] (g2),
      (c) -- [gluon, edge label' = $k_1$] (g1),
    };
  \end{feynman}
\end{tikzpicture}
\centering
\caption{Kinematics of double soft gluon emission, in the angular ordered parton shower, for ISR.}
\label{fig:kinematics2}
\end{figure}

To discuss the impact on the logarithmic accuracy of the recoil scheme, we focus now on the case of double soft gluon emission from an incoming quark line, as shown in Fig.~\ref{fig:kinematics2}.

Using the Sudakov decomposition described in the previous section:
\begin{subequations}
  \begin{eqnarray}
    p_3 &=& \alpha P;\\
    k_2 &=& (1-z_2)p_3 + \beta_2 n + p_{\perp 2}; \\
    p_2 &=& p_3-k_2 = z_2 p_3 -\beta_2 n - p_{\perp 2}; \\
    k_1 &=& (1-z_1)z_2 p_3 + \beta_1 n - (1-z_1)p_{\perp 2} + p_{\perp 1}; \\
    p_1 &=& p_2-k_1 = z_1z_2 p_3 -[\beta_1 + \beta_2]n - z_1 p_{\perp 2} - p_{\perp 1},     
\end{eqnarray}
\end{subequations}
where the $\beta_i$ are determined by the requirement that $k^2_i=0$ and $\alpha = x/(z_1 z_2)$. In a space-like shower backwards evolution is used, so the partons are labelled in ascending order the further they are from the hard process.

All recoil schemes are equivalent for the last emission, therefore
\begin{equation}
|p_{\perp2}| ^2 = (1-z_2)^2 \tilde{q}_2^2 = \epsilon_2^2 \tilde{q}_2^2,\, \qquad
p_2^2 = -(1-z_2) \tilde{q}_2^2= - \epsilon_2 \tilde{q}_2^2 ,
\label{eq:secondEmission}
\end{equation}
where we have introduced $\epsilon_2=1-z_2 \to 0$ in the soft limit.
The values of $p_1^2$ and $|p_{\perp1}|$ depend on the interpretation of the ordering variable.

\subsection{$p_\perp$-preserving scheme}
For the $p_{\perp}$-preserving scheme, we have
\begin{equation}
|p_{\perp1}| ^2 = (1-z_1)^2 \tilde{q}_1^2.
\end{equation}
Thus, from Eq.~\eqref{eq:virt_Generic} 
\begin{equation}
p_1^2 = - (1-z_1) \tilde{q}_1^2 -z_1(1-z_2)  \tilde{q}_2^2. 
\end{equation}
If both emissions are soft (\emph{i.e.}~ $1-z_i \equiv \epsilon_i \to 0$), we have
\begin{equation}
|p_{\perp1}| ^2 =  \epsilon_1^2 \tilde{q}_1^2, \qquad
p_1^2 = - \epsilon_1 \tilde{q}_1^2 - (1-\epsilon_1)\epsilon_2\tilde{q}_2^2 \approx - \epsilon_1 \tilde{q}_1^2 - \epsilon_2\tilde{q}_2^2.
\end{equation}
As for final-state radiation, this implies that the largest contribution to the virtuality can also come from subsequent emissions if for example $\epsilon_1 \ll \epsilon_2$.
However, since the transverse momentum of previously emitted gluons is unchanged, this scheme will reproduce the pattern of multiple independent soft emissions widely separated in angle.

\subsection{$q^2$-preserving scheme}
For the $q^2$-preserving scheme, we have
\begin{equation}
p_1^2 =  - (1-z_1) \tilde{q}_1^2,
\end{equation}
thus inverting Eq.~\eqref{eq:virt_Generic}
\begin{equation}
|p_{\perp1}|^2 =  (1-z_1)\left[(1-z_1) \tilde{q}_1^2 -z_1 (1-z_2) \tilde{q}_2^2 \right],
\end{equation}
which means that $|p_{\perp,1}|^2$ decreases and is not guaranteed to be positive.
 In the soft limit
\begin{equation}
|p_{\perp1}| ^2 \approx  \epsilon_1\left(\epsilon_1 \tilde{q}_1^2-\epsilon_2 \tilde{q}_2^2 \right), \qquad
p_1^2 = - \epsilon_1 \tilde{q}_1^2.
\end{equation}
Clearly if $\epsilon_2 \gg \epsilon_1$, the kinematics of the first emission is significantly modified. This, as we saw in Ref.~\cite{Bewick:2019rbu}, causes NLL issues, so we will not to consider this scheme further.

\subsection{Dot-product preserving scheme}
\label{sec:Dot_pTincrease}
When preserving the dot-product
\begin{equation}
p_1^2 = (p_2 -k_1)^2 = p_2^2 -2 p_2 \cdot k_1 =   -(1-z_2) \tilde{q}_2^2 -(1-z_1) \tilde{q}_1^2.
\end{equation}
From Eq.~\eqref{eq:ptDotGeneric} we have
\begin{align}
|p_{\perp1}|^2 = (1-z_1)^2 \left[ \tilde{q}_1^2 - p_2^2\right] = (1-z_1)^2 \left[ \tilde{q}_1^2+(1-z_2) \tilde{q}_2^2 \right]. 
\end{align}
Thus in the soft limit
\begin{equation}
|p_{\perp1}| ^2 =  \epsilon_1^2 (\tilde{q}_1^2+\epsilon_2 \tilde{q}_2^2)
\approx \epsilon_1^2 \tilde{q}_1^2, \quad
p_1^2 = - \epsilon_1 \tilde{q}_1^2 - \epsilon_2\tilde{q}_2^2.
\end{equation}
It is interesting to notice that in the case of ISR, subsequent emissions tend to increase $|p_{\perp1}| ^2$, while in the case of FSR the opposite behaviour takes place (see Ref.~\cite{Bewick:2019rbu}).
In any case, when $\epsilon_1$ and $\epsilon_2$ are both small, regardless of which one is smaller, this scheme reproduces the $p_\perp$-preserving one, thus it is capable of describing the matrix element for multiple independent soft gluon emissions.

\section{Global recoil strategy for $Z$ and $Z$+jet production}
\label{sec:global-rec-DY}
In a parton shower algorithm in which each parton showers independently, it is necessary to `re-assemble' the individual partons to produce the full event. Since the partons' virtualities are shifted by the showering, it is not possible to preserve all momentum components simultaneously, and some momentum must be shuffled between partons. We call the algorithm by which this is done the global recoil strategy.

In this section we summarize the global recoil strategy applied to the cases of $Z$ and $Z$+jet production. More details and the generalization to an arbitrary colour structure are discussed in appendix~\ref{sec:global_recoil}. 
We stress that in the presence of only soft and/or collinear emissions, the global recoil strategy amounts to power suppressed changes to the rapidity and the transverse momentum of partons emitted from initial state radiation and thus does not alter the discussion presented in the previous section.

Let us consider a Drell--Yan process, which at LO is described by the annihilation of a $q\bar{q}$ pair into a massive gauge boson.
Each incoming parton is identified as a \emph{shower progenitor} by the \Herwig{} angular-ordered shower, and thus showered independently. The two initial partons form a colour-connected neutral system so the transverse momentum of each new emission is defined with respect to the $q\bar{q}$ direction. After the showering phase, the original incoming partons have acquired a negative virtuality and a transverse momentum. The total transverse momentum imbalance is reabsorbed by the $Z$ boson, but there is some freedom in how to impose longitudinal momentum conservation. Three options are possible:
\begin{enumerate}
\item we fix the $Z$-boson rapidity;
\item we fix the longitudinal momentum of the $Z$ boson;
\item we preserve the new off-shell momentum of the shower progenitor that does not contain the hardest emission.
\end{enumerate}
The latter option is the default behaviour, as it allows for a simpler matching with higher order matrix elements. Indeed for one emission it exactly reproduces the kinematics of the Catani-Seymour initial-initial dipole~\cite{Catani:1996vz}. 
All these options ensure that, in the case of multiple soft emissions well separated in rapidity, the transverse momentum of the $Z$ is given by the vector sum of all the emissions' transverse momenta and that the $Z$ rapidity can receive only suppressed power corrections $\mathcal{O}\left(p_{\perp}^2/m_z^2\right)$.

In section Sec.~\ref{sec:generators} we will present our results for $Z$ and $Z$+jet matched predictions. In the latter case, also the final state quark can be identified as a \emph{shower progenitor}, which is colour connected to an initial parton. The global recoil strategy applied in this case is a hybrid between the one we just discussed for the production of a colour singlet, and the one developed for Deep Inelastic Scattering~(DIS) processes.
To be concrete, let us consider the $q e^- \to q e^-$ DIS process, which proceeds through a colour neutral $t$-channel exchange. The kinematic reconstruction and the transverse momentum are defined in the Breit frame, where the two quarks are back-to-back. Such a mapping must leave the momenta of the two electrons unchanged, thus the final-state quark and its \emph{children}, \emph{i.e.}~the partons produced during its parton-shower evolution, need to absorb the transverse momentum imbalance due to initial state radiation. A longitudinal boost is also applied to the incoming (outgoing) parton and its children to ensure that the $t$-channel propagator is preserved. 

For $Z$+jet production, the shower progenitor which leads to the hardest emission is reconstructed first together with its colour evolution partner. If they form an initial-initial dipole, the kinematic reconstruction devised for the Drell--Yan case is adopted, while if they form an initial-final dipole the one for DIS is implemented. Then one proceeds to the reconstruction of the remaining shower progenitor momentum, which is colour connected to a gluon jet which has already been reconstructed. In this case, the gluon jet will be boosted again to absorb the recoil of its second colour partner.

\section{Drell--Yan production}
\label{sec:generators}
To investigate the performance of the new recoil scheme for the
angular-ordered shower, we compared predictions for the DY process at
$\sqrt{s}=7,8$ and 13~TeV with ATLAS and CMS data.

We have considered several options for the treatment of the hardest
emission.  We simulated leading order~(LO) predictions, with matrix
element corrections~(MEC) to improve the description of the hardest
emission and to populate the \emph{dead zone}, \emph{i.e.}~the region
of phase space not populated by the angular-ordered shower. We also
simulated next-to-leading order~(NLO) predictions, obtained from
{\sf  Matchbox}~\cite{Platzer:2011bc} machinery, which allows the
inclusion of next-to-leading order corrections in either the {\sf 
MC@NLO}~\cite{Frixione:2002ik} or {\sf  POWHEG}~\cite{Nason:2004rx}
matching schemes.

For LO+MEC predictions, by default {\sf  Herwig} uses leading-order
parton distribution functions~(PDFs). However, the usage of NLO PDFs
is also possible as this introduces higher-order differences
beyond the level of accuracy of the calculation.\footnote{Here we
  refer to the PDF employed to evaluate the matrix element and to
  perform the ISR evolution. The underlying event~(UE) is not
  included, as it has no impact on the transverse momentum of the $Z$
  boson, which is only affected by ISR, in the \Herwig{}~{\sf 7} model. For
  the UE, the usage of LO PDFs is recommended as the combination of
  leading order matrix elements and PDFs is a better approximation to
  the NLO result than LO matrix elements and NLO PDFs due to the
  different behaviours of the gluon distribution at small $x$.} Thus,
for illustrative purposes, in the next sections we compare LO results
obtained with both LO and NLO PDFs, finding small differences as
expected.

When using the {\sf  POWHEG} method, it is possible to separate the
real emission contribution into a singular and a non-singular part,
and exponentiate in the Sudakov form factor only the former
contribution, while the latter is generated as a Born-like event with
a higher particle multiplicity. This separation is somewhat arbitrary
and in {\sf  Herwig} is controlled by the hard scale
profile~\cite{Bellm:2016rhh}, which we turn off in our simulation
(\emph{i.e.}~we exponentiate all the real corrections).

When the hardest emission generated by the {\sf  POWHEG}
algorithm is inside the phase space region accessible to the parton
shower algorithm (\emph{i.e.}~it is not in the \emph{dead zone}), the
kinematics are always reconstructed as a $Z$ event. However, in the case
that the hardest emission is in the \emph{dead zone}, we consider two
options. The first is to reconstruct the event as a
$Z$+jet hard process when the hardest emission is in the \emph{dead
zone}. We label this behaviour as ``{\sf  POWHEG}'' in the plots
below. But we also consider the option of always reconstructing the
event as a shower emission from a $Z$ event, which we label as ``{\sf 
POWHEG} (aS)''.
We consider the first to be more self-consistent, but the second is
actually the current {\sf  Herwig} default.
This choice has a negligible impact on the description of the small
$Z$-boson transverse momentum, but leads to sizeable differences in
the high-$p_{\perp}$ tails of distributions.

\subsection{Tuning}
\label{sec:tuning}
Before comparing {\sf  Herwig} predictions with the experimental data,
we need to tune the parameters sensitive to ISR: the intrinsic
$p_\perp$, \emph{i.e.}~the non-perturbative intrinsic transverse
momentum for the partons inside the incoming hadron, and the strong
coupling~$\alpha_s$ (which is given in the CMW~\cite{Catani:1990rr}
scheme). To this end, we have introduced the possibility of setting
the value of $\alpha_s$ for ISR independently from that which was
previously tuned for FSR. 
We do not consider an independent variation of the minimum transverse momentum used as a shower cutoff, $p_\perp^{\text{min}}$, because it is strongly anti-correlated with $\alpha_s$. Therefore it is sufficient only to tune $\alpha_s$ and set $p_\perp^{\text{min}}$ to its FSR value of $\approx 1$~GeV~ \cite{Bewick:2019rbu}\footnote{For the dot-product preserving scheme $p_\perp^{\text{min}}=0.958$~GeV, while for the $p_\perp$ preserving one $p_\perp^{\text{min}}=0.900$~GeV.}.
We consider only the dot-product and
transverse-momentum preserving schemes, as the virtuality preserving
scheme was found to have NLL issues in Ref.~\cite{Bewick:2019rbu}.

The parametric dependence of the distributions is obtained from
the\linebreak {\sf  Professor}~\cite{Buckley:2009bj} program, which
also finds the parameter values that minimise the $\chi^2$
distribution for the data we are using and parameters we are fitting.
To avoid being dominated only by points with a small experimental
uncertainty, during the minimization procedure we set the minimum
experimental uncertainty to 1\%.  We compare the events analysed with
{\sf  Rivet}~\cite{Buckley:2010ar} to the ATLAS Drell--Yan $Z$-boson
production data at 7~TeV~\cite{Aad:2014xaa,Aad:2012wfa,Aad:2011gj}.
To reduce correlations, we consider the transverse momentum of the
$Z$ boson reconstructed from muons, and the angular correlations
between $e^+e^-$ pairs produced from the $Z$ decay.

Since in our simulation we use only the matrix elements for $Z$ plus 0
or 1 jet, we are unable to perform a realistic description of the high
$p_\perp$ spectrum, which would require the matching with higher
multiplicity matrix elements as the parton shower approximation is not
valid in this region.  For this reason, when tuning we consider only
bins in which the $Z$ transverse momentum is smaller than 50~GeV and
$\phi^*_\eta<0.8$.\footnote{The variable $\phi^*_\eta$ is introduced
  in Ref.~\cite{Aad:2012wfa} and is defined as
\begin{equation}
\phi^*_\eta = \tan\left( \frac{\pi - \Delta\phi}{2} \right) \sin \theta^*,
\end{equation}
with $\Delta \phi$ being the azimuthal opening angle between the two
leptons and $\theta^*$ the scattering angle of the leptons with
respect to the proton beam direction in the rest frame of the dilepton
system.  }

The results of our tuning procedure for the several predictions
described in Sec.~\ref{sec:generators} are shown in
Tab.~\ref{tab:massless_parameters}.
We notice that the behaviour of the dot-product and $p_\perp$ preserving
schemes is similar, except when the {\sf  POWHEG} scheme is adopted and
the hardest emission inside the \emph{dead zone} is treated as a
shower emission (aS). For all the considered cases, the new
dot-product preserving scheme always yields a better chi-squared.
We also find that the value of $\alpha_s$ for ISR is considerably
larger than the one for FSR obtained in Ref.~\cite{Bewick:2019rbu}. In
particular, at LO (+MEC) the tuned value of $\alpha_s$ is very close
to 0.1256 in the CMW scheme, which corresponds to the well known value
$\alpha_s=0.118$ in the $\overline{\rm MS}$ scheme. The usage of NLO
PDFs for LO predictions does not yield a significant difference with
respect to the default choice of using LO PDFs.
On the other hand, when adopting the {\sf  MC@NLO} matching the value
of the strong coupling is always smaller and the $\chi^2$ is
slightly worse.
The predictions obtained with the {\sf  POWHEG} scheme are those with
the largest chi-squared, and the tuned value of $\alpha_s$ is always
larger than the expected value of 0.1256.  When we always treat
the hardest emission as a shower emission~(aS), the dot-product preserving scheme yields the best
chi-squared, with $\alpha_s=0.1255$. This is in contrast with our
expectations, as the treatment of emissions in the \emph{dead zone} as
part of the hard process is better 
motivated. Furthermore, as we have already said, this is the only case where
we find a significant discrepancy (both in the chi-squared and in the
tuned value of the strong coupling) between the $p_\perp$ and the
dot-product preserving schemes.

It is not a surprise that the value of the coupling obtained by tuning
the dot-product preserving scheme predictions is smaller than the one from the $p_\perp$ scheme, as
subsequent emissions tend to increase the transverse momentum of
previously emitted partons (see Sec.~\ref{sec:Dot_pTincrease}). This
behaviour is opposite to the FSR case, where instead the dot-product preserving scheme
yields an $\alpha_s$ value larger than in the $p_\perp$
scheme~\cite{Bewick:2019rbu}.

In {\sf POWHEG} we find a larger value of $\alpha_s$ since for the hardest emission, which is completely handled by {\sf Matchbox}, the CMW prescription is not included in the strong coupling evaluation. We have re-run the fits with the CMW prescription implemented and obtain $\alpha_s \approx 0.125$ with distributions and $\chi^2$ that are nearly identical.

\begin{table*}[tb]
\resizebox{\textwidth}{!}
{
\centering
 \begin{tabular}{|c||c|c|c|c|c||c|c|c|c|}
%  \begin{tabular}{|p{2.2cm}||p{3cm}|p{2cm}|p{2cm}|p{2cm}||p{2cm}|p{2cm}|p{2cm}|}
\hline
    Scheme  & \multicolumn{5}{|c||}{dot-product preserving}
            & \multicolumn{4}{|c|}{$p_\perp$ preserving}
    \\
 \hline   
 Accuracy   & LO (NLO PDF)  & LO & {\sf  MC@NLO} & {\sf  POWHEG} & {\sf  POWHEG} (aS) & LO & {\sf  MC@NLO} & {\sf  POWHEG}  & {\sf  POWHEG} (aS)\\
\hline
\multicolumn{10}{|c|}{ \bf ISR Shower Parameters} \\
\hline
\centering
$\alpha_s^{\rm ISR}$ & 0.1260 & 0.1247  & 0.1171 & 0.1341 & 0.1255  & 0.1264 & 0.1189 & 0.1352 & 0.1318  \\
\centering
Intrinsic $p_\perp$ & 0.984  & 1.008 & 1.780  & 1.552 & 1.803 & 0.865 & 1.679 & 1.542 & 1.696  \\
\hline
\multicolumn{10}{|c|}{ $\chi^2$ of best fit point } \\
\hline
\centering
$\chi^2$  & 521 & 590 & 719 & 1528 & 253 & 658 & 923 & 2551 & 914\\
\centering
$\chi^2$/NDOF & 2.8 & 3.2 & 3.9 & 8.3 & 1.4  & 3.6 & 5.0 & 13.8 & 5.0 \\
\hline
  \end{tabular}
 }
  \caption{Tuned parameters and $\chi^2$ for Drell--Yan $Z$-boson production events at 7~TeV .}
  \label{tab:massless_parameters}
\end{table*}

\subsection{Results}
In this section we present the results of our simulations of vector
boson production at the LHC and the impact of the recoil scheme and
matching procedure on the accuracy of these simulations.

\subsection*{$Z$ production at 7~TeV}

\begin{figure}[tb]
\includegraphics[width=0.5\textwidth]{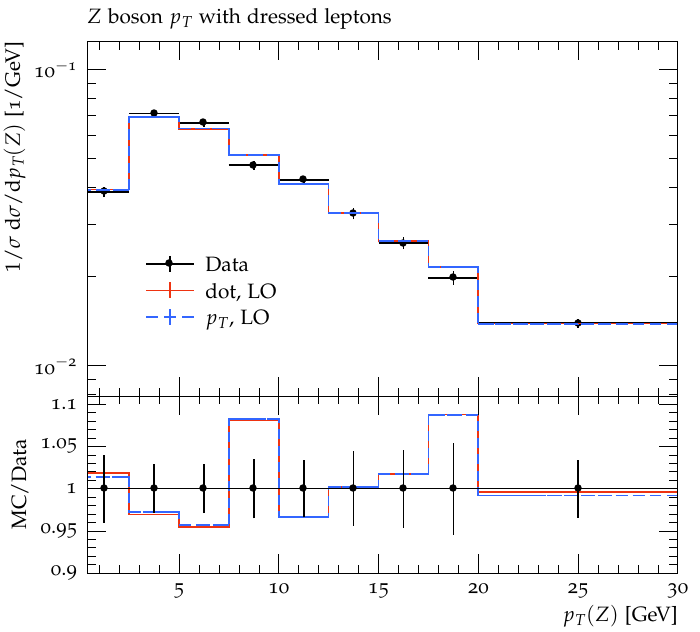}
\includegraphics[width=0.5\textwidth]{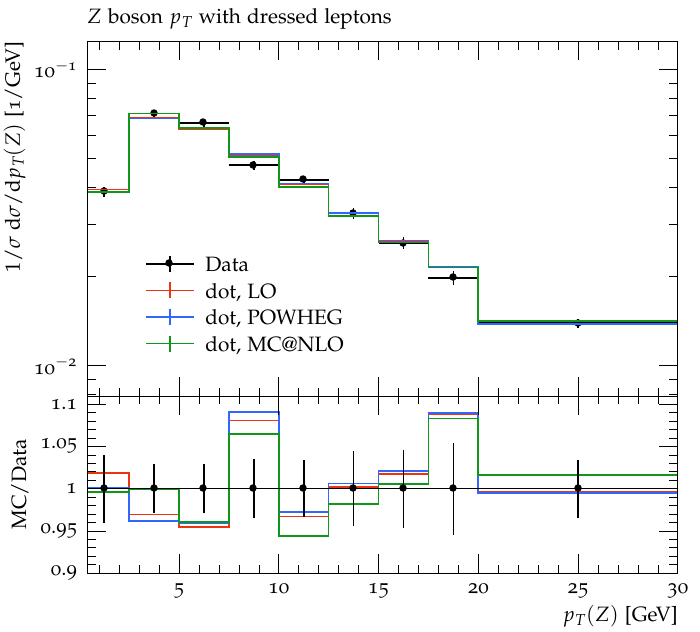}
\caption{Normalised differential cross section as a function of $Z$ $p_\perp$. CMS \cite{Chatrchyan:2011wt} data at 7~TeV compared to data generated by \Herwig{}. 
In the left pane, LO (plus MEC) predictions in the dot-product and $p_\perp$ preserving schemes, in the right pane LO and NLO predictions in the dot-product preserving scheme.}
\label{fig:Z-7TeV-pT}
\end{figure}
\begin{figure}[tb]
\centering
\includegraphics[width=0.49\textwidth]{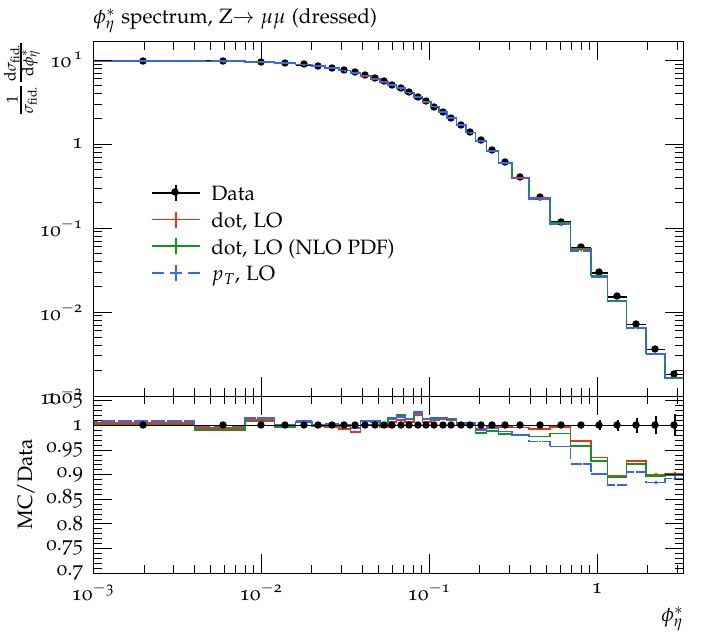}
\includegraphics[width=0.49\textwidth]{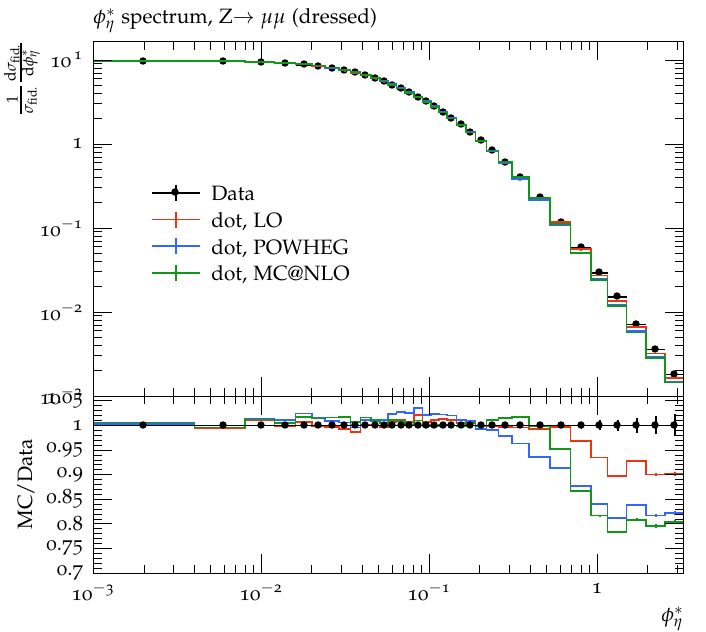}
\caption{Normalised differential cross section of $Z \to \mu^+ \mu^-$ as a function of the $\phi^*_\eta$ parameter of the $Z$ boson. \Herwig{} results compared to ATLAS \cite{Aad:2012wfa} data at 7~TeV. In the left pane, a comparison of LO results from the dot-product (with both LO and NLO PDFs) and $p_\perp$ preserving schemes. In the right pane, a comparison of LO and NLO results in the dot-product preserving scheme.}
\label{fig:Z-7TeV-phi}
\end{figure}
\begin{figure}[tb]
\includegraphics[width=0.5\textwidth]{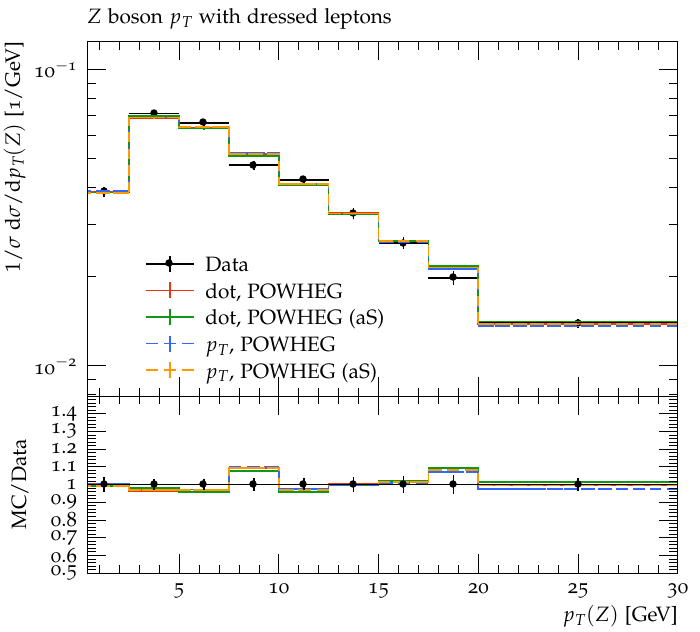}
\includegraphics[width=0.49\textwidth]{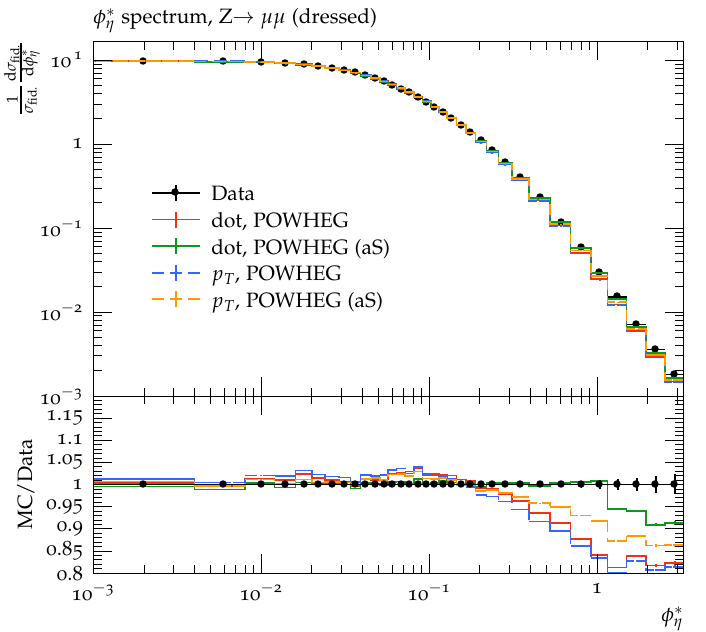}
\caption{Normalised differential cross section of $Z \to \mu^+ \mu^-$ as a function of $Z$ $p_\perp$~(left panel) and of the $\phi^*_\eta$ parameter~(right panel). \Herwig{} results, obtained using the {\sf  POWHEG} matching prescription, are compared to CMS \cite{Chatrchyan:2011wt} and ATLAS \cite{Aad:2012wfa} data at 7~TeV. }
\label{fig:Z-7TeV-powheg}
\end{figure}

We begin by illustrating the distributions of $Z$ boson transverse
momentum and $\phi_\eta^*$ at 7~TeV.  We expect to find good agreement
between all of our predictions and the data for small values of both
transverse momentum and $\phi^*_\eta$ as we have tuned our parameters
using the low-$p_{\perp}$ range of these distributions.

In Fig.~\ref{fig:Z-7TeV-pT} we show the CMS
measurement~\cite{Chatrchyan:2011wt} of the
transverse momentum of the $Z$ boson in the low $p_\perp$ region. We can observe that the dot-product and
$p_\perp$ preserving schemes are almost identical~(left pane),
and very small differences are found including higher order
corrections~(right pane).

Fig.~\ref{fig:Z-7TeV-phi} illustrates the distribution of the
$\phi_\eta^*$ parameter of the $Z$ boson measured by the ATLAS
collaboration\cite{Aad:2012wfa}. Like the transverse momentum results,
for small values of $\phi^*_\eta$ all the theoretical predictions are
very similar and agree well with data.  We observe deviations between
the recoil schemes~(left panel) and among the several matching
prescriptions~(right panel) only for larger values of
$\phi^*_\eta$. It is interesting to notice that the LO predictions
obtained using a NLO PDF~(green curve in the left plot) are very
similar to those obtained with the default LO PDF~(red curve). We
also notice that, at LO, the dot-product preserving scheme yields good agreement with
data up to $\phi^*_\eta \sim 0.8$, which is the upper value we used in
our tuning procedure, while $5\%$ differences with respect to the data
arise at $\phi^*_\eta \sim 0.8$ in the $p_\perp$-preserving scheme~(blue
curve in the left plot). Both the {\sf  MC@NLO} and {\sf  POWHEG} NLO predictions
(right plot) give a poorer description of the data, with small
discrepancies between $\phi_\eta^* \sim 0.02$ and $\phi_\eta^* \sim
0.1$, becoming very significant for $\phi_\eta^* > 0.2$ ({\sf  POWHEG}) or $0.5$ ({\sf  MC@NLO}).

In Fig.~\ref{fig:Z-7TeV-powheg} we compare the {\sf  POWHEG}
predictions obtained treating hardest emissions in the dead zone as
part of the hard process (\emph{i.e.}~as a genuine ``real emission''), with
those obtained treating the hardest emission always as a shower
emission (aS). We notice that for small values of the $Z$-boson $p_\perp$
(left panel), all the predictions are in good agreement. However,
large differences arise when looking at the hard tail of the
$\phi^*_\eta$ distribution (right panel), where all the \Herwig{}
predictions underestimate ATLAS data. The ``real-emission'' scheme,
which is the best theoretically motivated, leads to the largest
discrepancies with respect to the data.  The ``as shower''
treatment yields significantly different predictions between the dot-product
and $p_\perp$ preserving schemes.  As already seen in
Tab.~\ref{tab:massless_parameters}, {\sf  POWHEG} dot-product ``as
shower'' (aS) results~(green curve) are the ones that yield the best
description of the experimental data.

\subsection*{$W$ production at 7~TeV}
\begin{figure}[tb]
\centering
\includegraphics[width=0.49\textwidth]{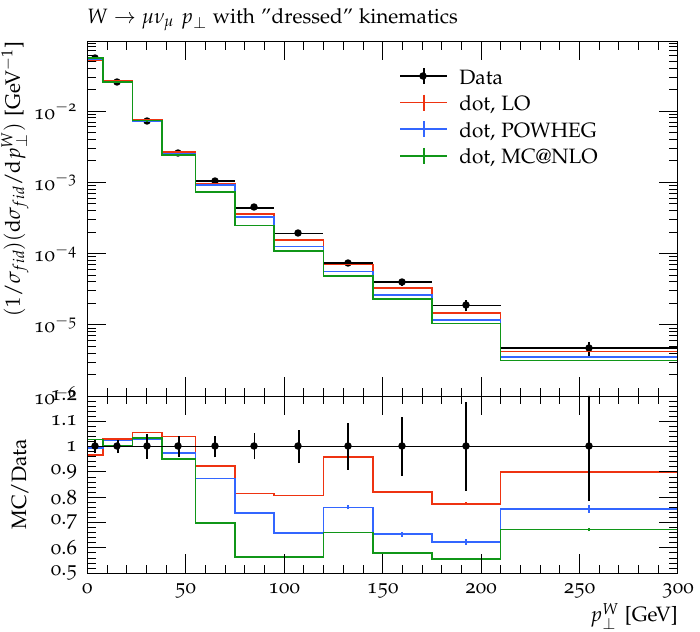}
\includegraphics[width=0.49\textwidth]{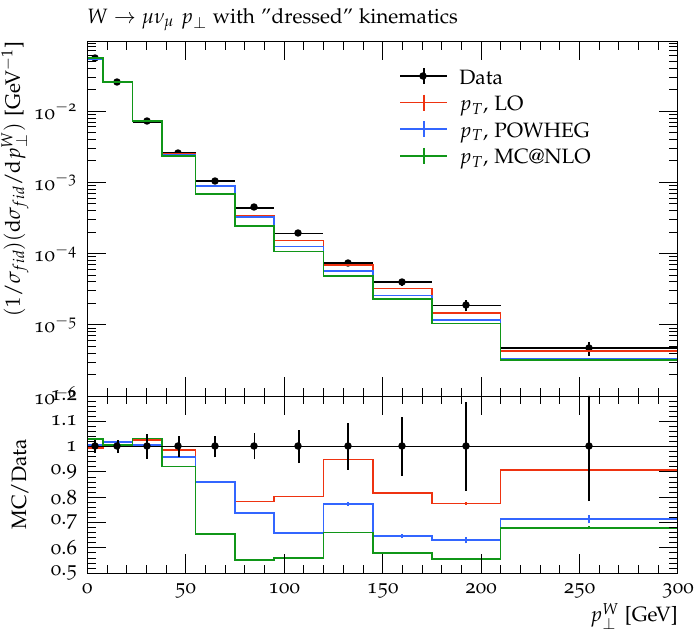}\\
\caption{Differential normalized cross section of $W \to \mu \nu_\mu$ as a function of $W$ $p_\perp$. \Herwig{} compared to ATLAS \cite{Aad:2011fp} data at 7~TeV.}
\label{fig:W-7TeV}
\end{figure}

As we have tuned using only the $Z$-boson transverse momentum
distribution, we can use the impact of our tune on the transverse
momentum of the $W$ boson to assess the universality of our tuning
procedure. In Fig.~\ref{fig:W-7TeV} we compare the ATLAS 7~TeV
measurement~\cite{Aad:2011fp} with our predictions in the dot-product~(left)
and $p_\perp$~(right) preserving schemes. For $W$-boson transverse momentum
values smaller that 50~GeV, all the predictions agree fairly well with
the data, which are however plagued by large uncertainties. For larger
values we see that all the theoretical predictions are systematically
lower than the data, particularly those obtained using the NLO matched simulations.

\subsection*{$Z$ production at 8~TeV}

\begin{figure}[tb]
\includegraphics[width=0.5\textwidth]{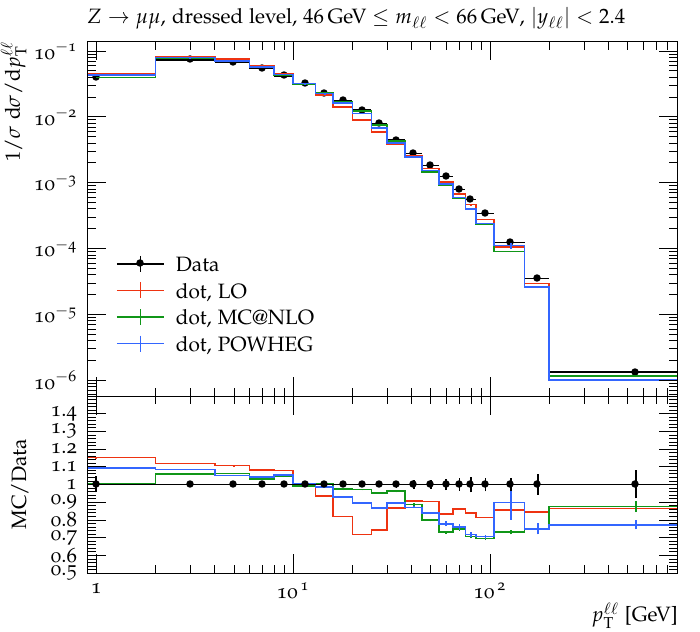}
\includegraphics[width=0.5\textwidth]{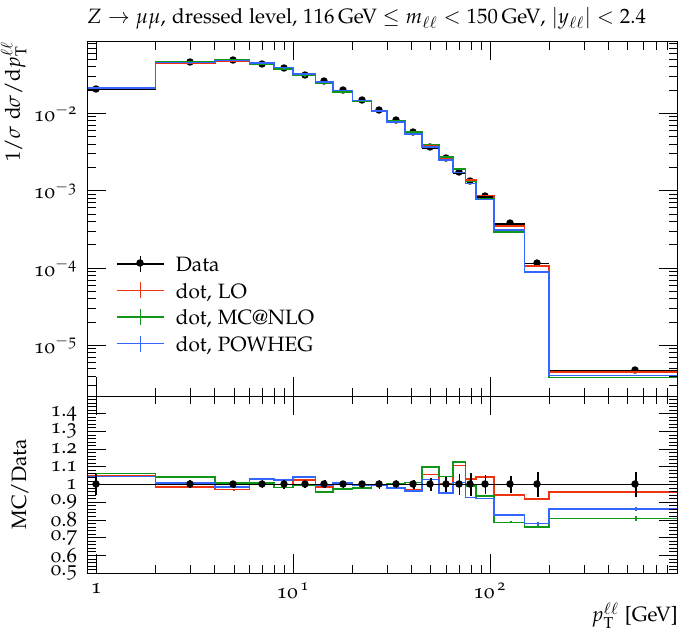}\\
\includegraphics[width=0.5\textwidth]{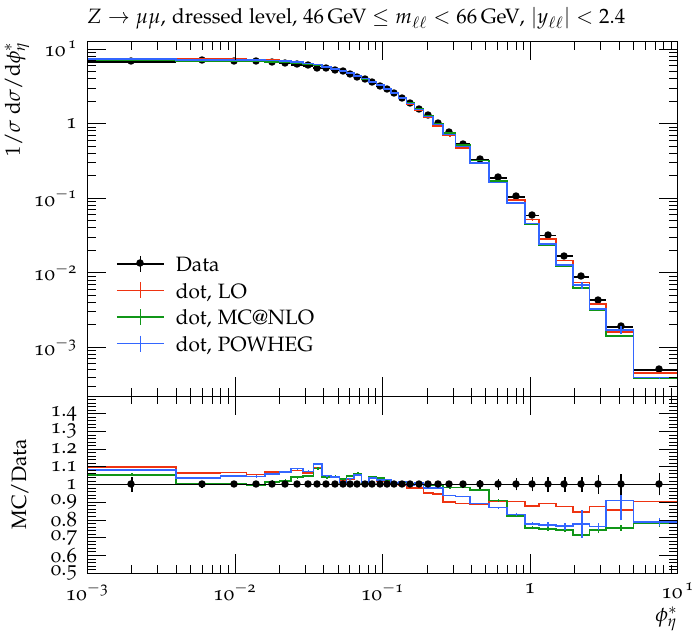}
\includegraphics[width=0.5\textwidth]{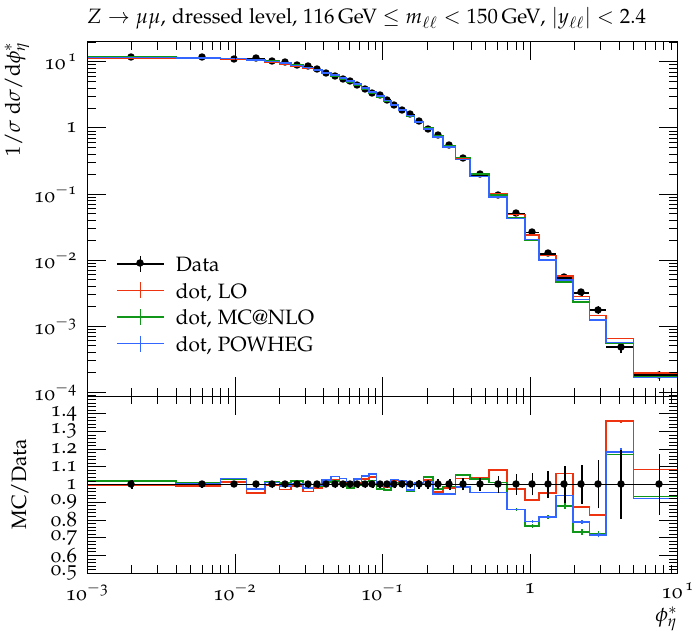}
\caption{Differential normalized cross section of $Z \to \mu^+ \mu^-$ as a function of 
  the $Z$ transverse momentum~(upper pane) and $\phi^*_\eta$~(low pane), below~(left) and above~(right) the $Z$ mass peak region. \Herwig{} compared ATLAS \cite{Aad:2015auj} at 8~TeV.}
\label{fig:Z-8TeV}
\end{figure}

We now examine the $Z$ boson distributions at 8~TeV, \emph{i.e.}~with
a centre-of-mass energy slightly above the one we adopted for the
tuning.  The ATLAS measurements of Ref.~\cite{Aad:2015auj} allow us to
investigate the behaviour below and above the $Z$-boson mass peak
region. Predictions in the peak region are similar to those described
in the previous section and not shown here.

In Fig.~\ref{fig:Z-8TeV} experimental data is compared to \Herwig{} results in the dot-product preserving scheme. Above the
$Z$-mass peak we get good agreement with the data for
$p_{\perp}<100$~GeV and $\phi^*_\eta<1$.  However, for masses below the
peak the LO predictions overpopulate the low-$p_\perp$
($\phi^*_\eta$) region and under populate the region with moderate
$p_\perp$ values. This feature is present also at NLO, but it is milder
and discrepancies are only of the order of a few percent.  In all
cases, the high-$p_\perp$ ($\phi^*_\eta$) tail is not properly
described. This is not surprising, as the proper treatment of this
region would require higher order matrix elements.

\subsection*{$Z$ production at 13~TeV}

\begin{figure}[h]
\includegraphics[width=0.5\textwidth]{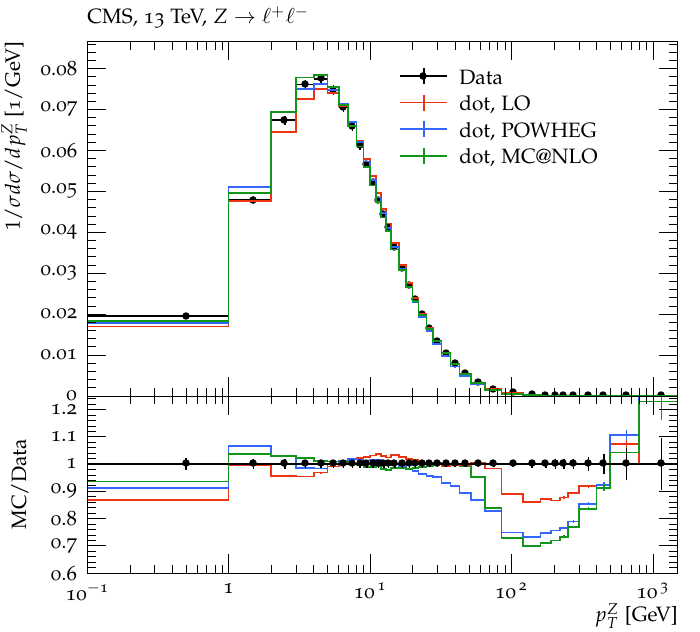}
\includegraphics[width=0.5\textwidth]{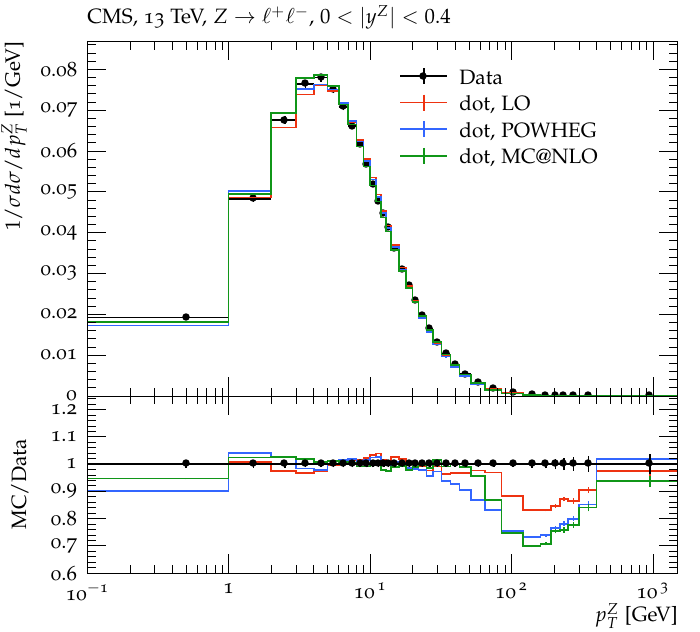}\\
\includegraphics[width=0.5\textwidth]{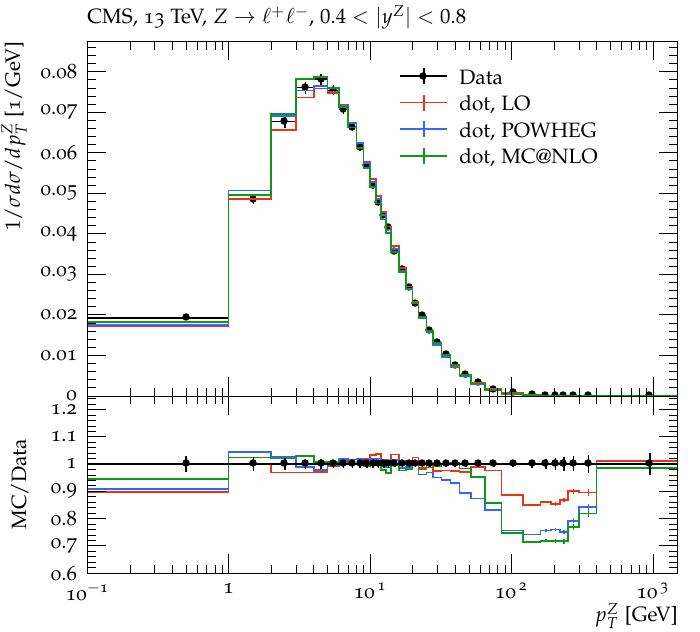}
\includegraphics[width=0.5\textwidth]{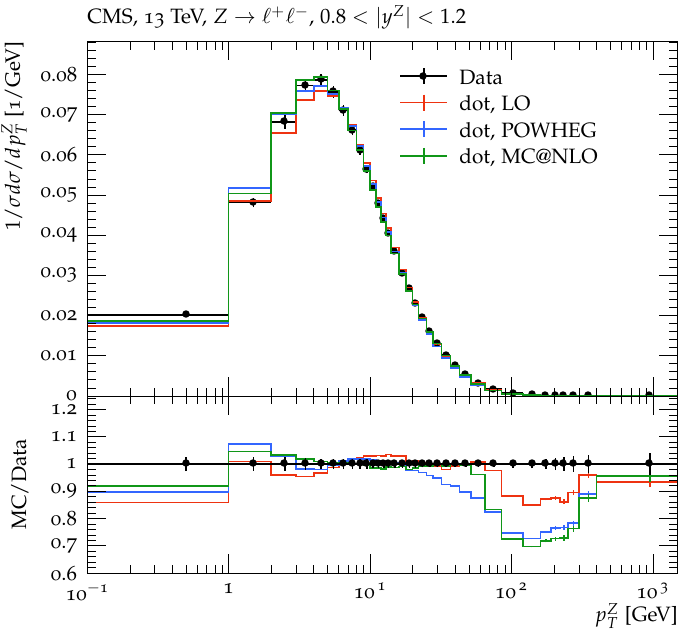}\\
\includegraphics[width=0.5\textwidth]{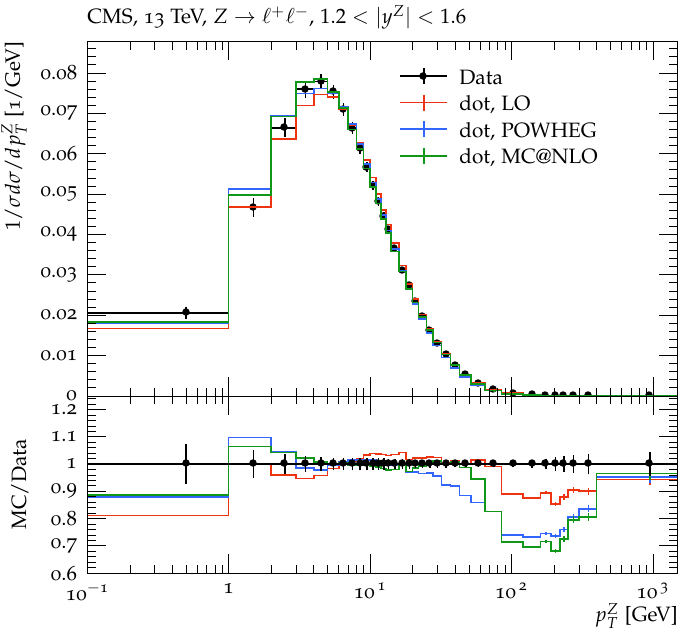}
\includegraphics[width=0.5\textwidth]{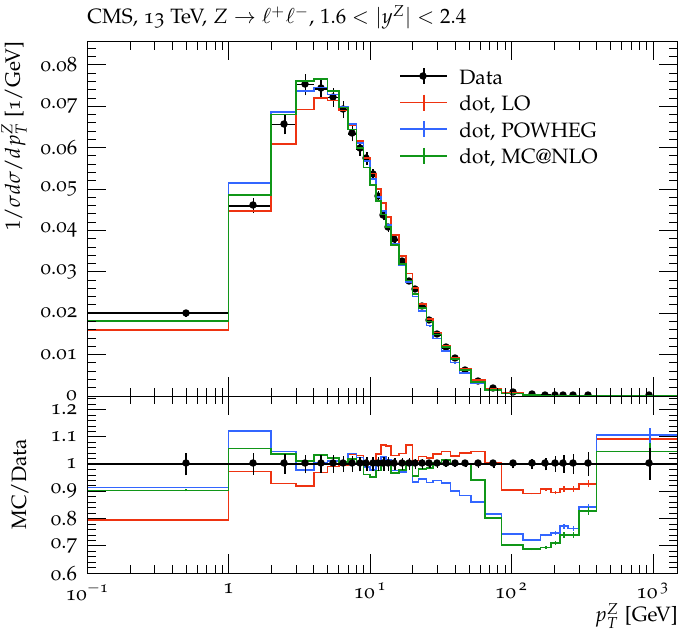}\\
\caption{Differential normalized cross section as a function of $Z$ $p_\perp$, broken down into different rapidity ranges.
\Herwig{} results are compared to CMS \cite{Sirunyan:2019bzr} data at 13~TeV.}
\label{fig:Z-13TeV}
\end{figure}

We conclude our phenomenological study by comparing \Herwig{}
predictions to CMS measurements at 13~TeV~\cite{Sirunyan:2019bzr}.
Fig.~\ref{fig:Z-13TeV} shows the distribution of the $Z$ boson
transverse momentum for several rapidity ranges, with the upper-left
plot showing the inclusive case. In Table~\ref{tab:chisq_13TeV} we
present the results of $\chi^2$ calculations for the full
distributions of Z boson production at 13 TeV: $p_\perp$,
$\phi^*_\eta$ and $y$ ($Z$ boson rapidity). The table also shows the total $\chi^2$ of the
$p_\perp$ distributions for the $y$ sub-ranges. For this calculation
we consider only $p_\perp<50$ and $\phi^*_\eta<0.8$.

\clearpage

We notice that all predictions, particularly those obtained using
NLO matching prescriptions, track the data fairly well around the peak
region, $p_{\perp}^{Z}\approx 10$~GeV, however large differences are found
both in the large and in the small $p_\perp$ regions.
For very small $p_{\perp}^Z$ values, the distribution is very
  sensitive to the modelling of the intrinsic $p_\perp$, which we have
  tuned at $7$~TeV. For an improved description of the data, a new
  tuning could be performed for each centre-of-mass
  energy. Alternatively, the model of \cite{Gieseke:2007ad} could be
  used to predict the amount of transverse momentum built up by
  non-perturbative smearing throughout the perturbative evolution, as
  a function of the partonic and hadronic centre-of-mass energies.

\begin{table}[t]
\resizebox{\textwidth}{!}
{
  \begin{tabular}{|c|c|c|c|c|c|c|}
\hline
    Scheme  & dot LO & dot {\sf  MC@NLO} & dot {\sf  POWHEG} & $p_\perp$ LO & $p_\perp$ {\sf  MC@NLO} & $p_\perp$ {\sf  POWHEG} \\
\hline
\multicolumn{7}{|c|}{ $\chi^2$ } \\
\hline
$p_\perp$ & 57.94 & 30.81 & 300.2 & 130.8 & 37.75 & 517.1 \\
$\phi^*_\eta$ & 36.30 & 683.2 & 1230 & 243.3 & 1164 & 1712 \\
$y$ & 106.4 & 22.84 & 11.60 & 102.4 & 24.35 & 11.16 \\
$p_\perp$, $y$ bins & 401.8 & 147.8 & 1355 & 729.8 & 171.8 & 2332 \\
\hline
\hline
\multicolumn{7}{|c|}{ Reduced $\chi^2$ } \\
\hline
$p_\perp$ ($\chi^2$/24) & 2.414 & 1.284 & 12.51 & 5.448 & 1.573 & 21.55 \\
$\phi^*_\eta$ ($\chi^2$/26) & 1.396 & 26.90 & 47.30 & 9.359 & 44.76 & 65.86 \\
$y$ ($\chi^2$/12) & 8.863 & 1.903 & 0.9668 & 8.535 & 2.029 & 0.9298 \\
$p_\perp$, $y$ bins ($\chi^2$/120) & 3.348 & 1.232 & 11.29 & 6.081 & 1.431 & 19.43 \\
\hline
  \end{tabular}
 }
  \caption{$\chi^2$ results for $Z$ Drell--Yan distributions at 13~TeV 
  compared to CMS data\cite{Sirunyan:2019bzr}. For the $Z$-boson transverse momentum distributions, the $\chi^2$ is computed only in the region $p_\perp < 50$~GeV, while for $\phi^*_\eta$ we consider only $\phi^*_\eta<0.8$. The ``$p_\perp$, $y$ bins'' rows refer to the $p_\perp$ distributions broken up into 5 distributions based on $y$.
   }  
  \label{tab:chisq_13TeV}
\end{table}

Furthermore, from Table~\ref{tab:chisq_13TeV}, we notice that given a
matching procedure, the results obtained in the dot-product preserving
scheme always yield a smaller $\chi^2$. The {\sf  POWHEG} predictions
always lead to the worst descriptions of the data, apart from the
rapidity distribution of the vector boson ($y$).  This is not
unexpected, since the parton shower is not allowed to change the
rapidity of the $Z$ boson\footnote{It should be noted that with the current reconstruction scheme $y$ is not preserved, but it is subject only to minor changes as it is  non-zero at leading-order.}, thus the $\chi^2$ value associated with the
$y$ distributions reported in Table~\ref{tab:chisq_13TeV} depends only
on the matching procedure, and not on the recoil scheme.

\section{Conclusions}
\label{sec:concl}
In this paper we have generalized our previous
study~\cite{Bewick:2019rbu} of the logarithmic accuracy of
angular-ordered parton showers to include initial-state radiation.

Similarly to the final-state case, we find that the $q^2$-preserving scheme has
problems in the strongly-ordered regime, with later emissions being
able to modify the kinematics of an earlier soft emission and diminish
its logarithmic accuracy. For this reason, we have not considered it
further.

We have extended the definition of the dot-product preserving scheme
introduced in Ref.~\cite{Bewick:2019rbu} to ISR.
The implementation of this recoil option thus enables a
consistent treatment of radiation which is emitted from the initial and final states.
Like the $p_\perp$-preserving scheme, this does not modify earlier kinematics
significantly, so that multiple-gluon emission inherits the correct
radiation pattern of single-gluon emission. One significant difference
compared to the final-state case is that in the initial state,
additional soft gluon emissions add to the transverse momentum
generated by the first emission, while in the final-state case, they
reduce it. This has an impact on the value of $\alpha_s$ fitted to
data, which we allow to be different for ISR and FSR.

Based on their logarithmic accuracy, we conclude that either the
$p_\perp$-preserving or dot-product preserving scheme can be used
for ISR. For the sake of consistency between the ISR case and the
FSR case, where the dot-product preserving scheme is recommended, we
recommend it also for ISR.

We have performed a dedicated tune of the non-perturbative parameters using $Z$-production data at 7 TeV for each recoil scheme and matching procedure that we considered.
Due to the importance of gauge boson transverse momentum distributions for the LHC physics program, we have performed a phenomenological study of the Drell-Yan process using these tuned showers. Since the description of data is better with the
dot-product preserving scheme than with the $p_\perp$-preserving
scheme (regardless of the matching prescription employed), we recommend the use of the dot-product preserving scheme for the additional reason that it provides a better modelling of the data.

It is impossible to consider the description of data without also
considering the choice of hard process and higher order matching
scheme. Of the NLO schemes, the {\sf  POWHEG}
(aS) scheme with the dot-product preserving shower gives the best
description of data. We consider this scheme to be somewhat
inconsistent, because it treats real emission in the dead zone that
the shower cannot populate, as if it was a parton shower
emission. Nevertheless, because of its significantly better
description of data, we recommend that it remain \Herwig's default
setting for now. The dot-product preserving reconstruction scheme will
become the default also for ISR from the next \Herwig{} release.

\acknowledgments

We thank our fellow \Herwig{} authors for useful
discussions, in particular Simon Pl\"{a}tzer for useful comments on the manuscript.
This work has received funding from the UK Science and Technology Facilities Council
(grant numbers ST/P000800/1, ST/P001246/1, ST/T001011/1, ST/T001038/1) and the European Union’s Horizon 2020 research and innovation 
programme as part of the Marie Sk\l{}odowska-Curie Innovative Training Network MCnetITN3
(grant agreement no. 722104).
S.F.R.’s work was also supported by the European Research Council (ERC) under the European Union’s Horizon 2020 research and innovation programme (grant agreement No. 788223, PanScales).
G.B. thanks the UK Science and Technology
Facilities Council for the award of a studentship.

\appendix

\section{Global recoil}
\label{sec:global_recoil}

In this section we briefly describe the global recoil that is applied to hadron-collider events at the end of the showering phase to achieve momentum conservation. 
An exhaustive description of the recoil strategy of the previous \Herwig{}{\sf  ++} generator, which shares many similarities with the current one (adopted since the {\sf  7.0} release), can be found in Sec.~6.5.2 of Ref.~\cite{Bahr:2008pv}. 

After the parton shower evolution, the space-like shower progenitors (\emph{i.e.}~the partons colliding in the hard process) and the time-like ones (\emph{i.e.}~the final-state partons arising from the hard scattering) are no longer on their mass shell, having acquired a negative or positive virtuality. Furthermore, the colliding particles have also acquired some transverse momentum that must be redistributed among the final-state progenitors and their daughter partons.
We therefore need to perform some momentum reshuffling to ensure momentum conservation.
How this is performed depends on whether the colour partner is an initial- or final-state parton.

The details of the algorithm for final-final correlations can be found in Sec.~6.4.2 of  Ref.~\cite{Bahr:2008pv}, here we focus on the case where ISR is involved.

\subsection{Drell--Yan: initial-initial correlations}

When we consider the production of colour-singlet systems, such as electroweak gauge bosons in Drell--Yan processes, we only have an initial-initial dipole. 

We use the hadronic beam momenta $p_{\oplus}$ and $p_{\ominus}$ to define the Sudakov basis for the initial-shower algorithms.
The suffix $\oplus$ denotes the particle incident from the $+z$ direction, while $\ominus$ from the $-z$ direction.
The momenta of the colliding partons $q_{\oplus}$ and $q_{\ominus}$ can be written
\begin{equation}
q_{\opm} = \alpha_{\opm} p_{\opm} + \beta_{\opm} p_{\omp} + q_{\perp\opm},
\end{equation}
where $q_{\perp\oplus}$ and $q_{\perp\ominus}$ are two space-like vectors orthogonal to the beam momenta.

We denote by $p_{\rm \scriptsize cms}$ the original final state
momentum, \emph{i.e.}~the momentum of the colour singlet system. Prior
to the inclusion of the shower 
\begin{equation}
p_{\rm \scriptsize cms} = x_\oplus p_\oplus + x_\ominus p_\ominus,
\end{equation}
but now the sum of the momenta of the incoming shower progenitors, $q_{\rm \scriptsize cms} = q_\oplus + q_\ominus$ is different from 
$p_{\rm \scriptsize cms}$. We can thus introduce two rescaling factors, $k_{\opm}$ to define the shuffled momenta $q^\prime$
\begin{equation}
q^\prime_{\opm} = \alpha_{\opm} k_{\opm} p_{\opm} + \frac{\beta_{\opm}}{k_{\opm}} p_{\omp} + q_{\perp\opm}, 
\end{equation}
which satisfy $q^{2\prime}_{\opm}=q^2_{\opm}$, and
\begin{equation}
q^\prime_{\rm cms} =q^\prime_{\oplus}+q^\prime_{\ominus} = 
\left(\alpha_{\oplus} k_{\oplus} + \frac{\beta_{\ominus}}{k_{\ominus}}\right)
p_{\oplus} + \left(\alpha_{\ominus} k_{\ominus} + \frac{\beta_{\oplus}}{k_{\oplus}}\right)
p_{\ominus} + q_{\perp \oplus}+q_{\perp \ominus}
\end{equation}
By imposing $q^{\prime2}_{\rm cms}=p^2_{\rm \scriptsize cms} = x_{\oplus} x_{\ominus} s$, where $\sqrt{s}$ in the centre-of-mass energy of the hadronic collision, we obtain a constraint on the product of the rescaling factors $k_{\oplus\ominus}=k_{\oplus}k_{\ominus}$:
\begin{equation}
\alpha_{\oplus}\alpha_{\ominus}k_{\oplus\ominus}^2
+\left(\alpha_{\oplus}\beta_{\ominus}+\alpha_{\ominus}\beta_{\oplus}-x_{\oplus} x_{\ominus} +\frac{(q_{\perp \oplus}+q_{\perp \ominus})^2}{s}
\right)k_{\oplus\ominus} + \beta_{\oplus}\beta_{\ominus} =0.
\label{eq:prodrescaling}
\end{equation}
It is trivial to check that if no emission has occurred, \emph{i.e.}~$\alpha_{\opm}=x_{\opm}$, $\beta_{\opm}=0$ and $q_{\perp \opm}=0$, then $k_{\oplus}=k_{\ominus}=1$ is a solution of eq.~\eqref{eq:prodrescaling}. 

By default, we set the rescaling factor of the progenitor which had the largest transverse momentum emission to $k_{\oplus\ominus}$, and the other rescaling factor to 1. This choice makes matching with higher order matrix elements simpler as for one emission it exactly reproduces the kinematics of the Catani-Seymour dipole~\cite{Catani:1996vz}.\footnote{An alternative option, which was the default in \Herwig{}{\sf  ++} and {\tt FORTRAN HERWIG}, is to preserve the rapidity of the colour-singlet system, \emph{i.e.}~that the ratio of $p_{\oplus}$ and $p_{\ominus}$ is identical in $q^\prime_{\rm cms}$ and $p^{\rm cms}$:
\begin{equation}
k_{\oplus}^2 = k_{\oplus\ominus} \frac{x_{\oplus}\left(\beta_{\oplus}+\alpha_{\ominus}k_{\oplus\ominus}\right)}{x_{\ominus}\left(\beta_{\ominus}+\alpha_{\oplus}k_{\oplus\ominus}\right)}.
\end{equation}
The last option is to preserve the longitudinal momentum, which leads to
\begin{equation}
\left( \alpha_{\oplus}+\frac{\beta_{\ominus}}{k_{\oplus\ominus}}\right)k_{\oplus}^2+(x_{\oplus}-x_{\ominus})k_{\oplus}-(\alpha_{\ominus}k_{\oplus\ominus}+\beta_{\oplus})=0.
\end{equation}
}

Since $q_{\opm}$ and $q^\prime_{\opm}$ have the same virtuality, it is possible to define a boost to transform $q_{\opm}$ to $q^\prime_{\opm}$. This boost is then applied to all the time-like children and the final space-like child produced during the showering phase.

We then need a second boost, which is applied to the original colour singlet final state, from $p_{\rm cms}$ to $q^\prime_{\rm cms}$, in order to absorb the transverse momentum $q_{\perp \oplus}+q_{\perp \ominus}$ that the colliding partons have acquired.

It is easy to check that
\begin{equation}
k_{\oplus\ominus} = 1 + \mathcal{O}\left(\frac{q^2_{\oplus}}{s}\right)+ \mathcal{O}\left(\frac{q^2_{\ominus}}{s}\right),
\end{equation}
 \emph{i.e.}~the rescaling coefficients are equal to 1 plus power-suppressed corrections, thus the boosts applied to the daughters of the time-like shower progenitors do not alter the logarithmic accuracy of the result.\footnote{By default we boost only the time-like jet that contains the hardest emission, however if we adopt one of the alternative reconstruction options we need to build two separate boosts, one for each incoming shower progenitor.}

\subsection{Deep inelastic scattering: initial-final correlations}

We now consider deep inelastic processes, \emph{i.e.}~when the incoming parton with momentum $p_{\rm in}$ is colour connected to an outgoing parton with momentum $p_{\rm out}$. We want our recoil strategy to preserve the transferred momentum, defined as
\begin{equation}
Q^2 = - (p_{\rm in}-p_{\rm out})^2.
\end{equation}
In the Breit frame
\begin{align}
p_{\rm in} =& \frac{Q}{2}\left[ 1+ \frac{m_{\rm out}^2}{Q^2};\, \vec{0}, +1+ \frac{m_{\rm out}^2}{Q^2} \right] \\
p_{\rm out} =& \frac{Q}{2}\left[ 1+ \frac{m_{\rm out}^2}{Q^2};\, \vec{0}, -1+\frac{m_{\rm out}^2}{Q^2} \right] \\
\Delta p =& p_{\rm in}-p_{\rm out} = Q \left[ 0; \, \vec{0}, 1 \right],
\end{align}
where $m_{\rm out}$ is the on-shell mass of the outgoing shower progenitor colour connected to the incoming one.
We introduce a set of basis vectors
\begin{equation}
n_1 = Q \left[1; \vec{0}, 1 \right], \qquad n_2 = Q \left[1; \vec{0}, -1 \right],
\end{equation}
so that 
\begin{equation}
\Delta p = \frac{1}{2}\left(n_1-n_2\right),
\end{equation}
and the momentum of the incoming jet, after the radiation, can be written as
\begin{equation}
q_{\rm in} = \alpha_{\rm in} n_1 + \beta_{\rm in} n_2 + q_{\perp}.
\end{equation}
The transverse momentum component of $q_{\rm in}$ must be absorbed by the outgoing progenitor $q_{\rm out}$ (and its daughters), so we first perform a rotation that leads to
\begin{equation}
q_{\rm out} = \alpha_{\rm out} n_1 + \beta_{\rm out} n_2 + q_{\perp},
\end{equation}
where $\beta_{\rm out} = \frac{1}{2}$.

We introduce rescaling factors $k_{\rm in,out}$ that allow us to define the shuffled momenta
\begin{equation}
q^\prime_{\rm in,out} = \alpha_{\rm in, out} k_{\rm in, out}n_1 + \frac{\beta_{\rm in, out}}{k_{\rm in, out}} n_2 + q_{\perp}.
\end{equation}
We impose 
\begin{equation}
\Delta p = p_{\rm in} -p_{\rm out} = q^\prime_{\rm in} -q^\prime_{\rm out},
\end{equation}
which leads to 
\begin{equation}
\alpha_{\rm in} k_{\rm in} - \alpha_{\rm out} k_{\rm out} =\frac{1}{2}, \qquad \frac{\beta_{\rm in}}{k_{\rm in}}- \frac{\beta_{\rm out}}{k_{\rm out}}= -\frac{1}{2}.
\end{equation}
Each of these rescalings can be implemented via a boost applied to the progenitor and to its daughter particles.

If we assume that $m_{\rm out} \ll Q^2$, we can write
\begin{equation}
\alpha_{\rm in} = \beta_{\rm out}= \frac{1}{2},\qquad \beta_{\rm in} = \mathcal{O}\left(\frac{p^2_{\rm in}}{Q} \right), \qquad  \alpha_{\rm out} = \mathcal{O}\left(\frac{p^2_{\rm out}}{Q} \right),
\end{equation}
which implies that
\begin{equation}
k_{\rm in,out} = 1 + \mathcal{O}\left(\frac{p^2_{\rm in}}{Q}\right)+ \mathcal{O}\left(\frac{p^2_{\rm out}}{Q} \right),
\end{equation}
\emph{i.e.}~the boost only leads to power-suppressed corrections and does not alter the logarithmic structure of the result.

\subsection{General case}
For more complicated colour structures (like \emph{e.g.}~$Z$+jet production) we need a more general procedure.

The default approach used by \Herwig{}{\sf  ++} from version 2.3  and the one employed by {\tt  FORTRAN HERWIG} and \Herwig{}{\sf  ++} versions prior to 2.3 are both detailed in Ref.~\cite{Bahr:2008pv},  here we want to present the new approach introduced in \Herwig{}~{\sf  7}, which uses the information on the colour structure as much as possible.

The jet associated with the progenitor which leads to the hardest emission is reconstructed first. By default, its evolution partner is also reconstructed.\footnote{Optionally, the evolution partner absorbs only the recoil, therefore staying on its mass-shell, but the full jet arising from its evolution is not reconstructed yet. }
The procedure is then repeated with the next unreconstructed jet with the hardest emission. 
Since a gluon has two colour partners, it is shifted twice, once by
the recoil from each of its partners.  

For the case of DY production with matrix element corrections, the
hardest emission is considered as part of the hard process (and thus
treated as a shower progenitor and not as a shower emission) only when it is inside the dead zone.
For {\sf  POWHEG} matched DY production, by default a profile function is employed to decide whether the first emission should be treated as a shower emission (and exponentiated in the Sudakov) or as part of the hard process.
However, in this work, we switch off this profiling mechanism and we always exponentiate the hardest emission, but we treat it as a shower progenitor when it is inside the \emph{dead zone}.
For {\sf  MC@NLO} matched DY production, the first emitted parton is always interpreted as part of the hard process, \emph{i.e.}~as a shower progenitor. 

In any of these cases, the global recoil is obtained by rescaling the momenta of the progenitors. However, since such rescalings are equal to $1$ plus power-suppressed corrections, they do not interfere with the logarithmic structure of the result.

\providecommand{\href}[2]{#2}\begingroup\raggedright\endgroup


\begin{thebibliography}{10}

\bibitem{Dasgupta:2018nvj}
M.~Dasgupta, F.~A. Dreyer, K.~Hamilton, P.~F. Monni and G.~P. Salam,
  \emph{{Logarithmic accuracy of parton showers: a fixed-order study}},
  \href{https://doi.org/10.1007/JHEP09(2018)033}{\emph{JHEP} {\bfseries 09}
  (2018) 033} [\href{https://arxiv.org/abs/1805.09327}{{\ttfamily
  1805.09327}}].

\bibitem{Sjostrand:2004ef}
T.~Sjostrand and P.~Z. Skands, \emph{{Transverse-momentum-ordered showers and
  interleaved multiple interactions}},
  \href{https://doi.org/10.1140/epjc/s2004-02084-y}{\emph{Eur. Phys. J.}
  {\bfseries C39} (2005) 129}
  [\href{https://arxiv.org/abs/hep-ph/0408302}{{\ttfamily hep-ph/0408302}}].

\bibitem{Hoche:2015sya}
S.~H\"{o}che and S.~Prestel, \emph{{The midpoint between dipole and parton
  showers}}, \href{https://doi.org/10.1140/epjc/s10052-015-3684-2}{\emph{Eur.
  Phys. J.} {\bfseries C75} (2015) 461}
  [\href{https://arxiv.org/abs/1506.05057}{{\ttfamily 1506.05057}}].

\bibitem{Bewick:2019rbu}
G.~Bewick, S.~Ferrario~Ravasio, P.~Richardson and M.~H. Seymour,
  \emph{{Logarithmic accuracy of angular-ordered parton showers}},
  \href{https://doi.org/10.1007/JHEP04(2020)019}{\emph{JHEP} {\bfseries 04}
  (2020) 019} [\href{https://arxiv.org/abs/1904.11866}{{\ttfamily
  1904.11866}}].

\bibitem{Catani:1990rr}
S.~Catani, B.~R. Webber and G.~Marchesini, \emph{{QCD} {C}oherent {B}ranching
  and {S}emi-{I}nclusive {P}rocesses at {L}arge x}, {\emph{Nucl. Phys.}
  {\bfseries B349} (1991) 635}.

\bibitem{Platzer:2009jq}
S.~Platzer and S.~Gieseke, \emph{{Coherent Parton Showers with Local Recoils}},
  \href{https://doi.org/10.1007/JHEP01(2011)024}{\emph{JHEP} {\bfseries 01}
  (2011) 024} [\href{https://arxiv.org/abs/0909.5593}{{\ttfamily 0909.5593}}].

\bibitem{Catani:1996vz}
S.~Catani and M.~Seymour, \emph{{A General algorithm for calculating jet
  cross-sections in NLO QCD}},
  \href{https://doi.org/10.1016/S0550-3213(96)00589-5}{\emph{Nucl. Phys. B}
  {\bfseries 485} (1997) 291}
  [\href{https://arxiv.org/abs/hep-ph/9605323}{{\ttfamily hep-ph/9605323}}].

\bibitem{Platzer:2011bc}
S.~Platzer and S.~Gieseke, \emph{{Dipole Showers and Automated NLO Matching in
  Herwig++}}, \href{https://doi.org/10.1140/epjc/s10052-012-2187-7}{\emph{Eur.
  Phys. J. C} {\bfseries 72} (2012) 2187}
  [\href{https://arxiv.org/abs/1109.6256}{{\ttfamily 1109.6256}}].

\bibitem{Frixione:2002ik}
S.~Frixione and B.~R. Webber, \emph{{Matching NLO QCD computations and parton
  shower simulations}},
  \href{https://doi.org/10.1088/1126-6708/2002/06/029}{\emph{JHEP} {\bfseries
  06} (2002) 029} [\href{https://arxiv.org/abs/hep-ph/0204244}{{\ttfamily
  hep-ph/0204244}}].

\bibitem{Nason:2004rx}
P.~Nason, \emph{{A New method for combining NLO QCD with shower Monte Carlo
  algorithms}},
  \href{https://doi.org/10.1088/1126-6708/2004/11/040}{\emph{JHEP} {\bfseries
  11} (2004) 040} [\href{https://arxiv.org/abs/hep-ph/0409146}{{\ttfamily
  hep-ph/0409146}}].

\bibitem{Bellm:2016rhh}
J.~Bellm, G.~Nail, S.~Plätzer, P.~Schichtel and A.~Siódmok, \emph{{Parton
  Shower Uncertainties with Herwig 7: Benchmarks at Leading Order}},
  \href{https://doi.org/10.1140/epjc/s10052-016-4506-x}{\emph{Eur. Phys. J. C}
  {\bfseries 76} (2016) 665}
  [\href{https://arxiv.org/abs/1605.01338}{{\ttfamily 1605.01338}}].

\bibitem{Buckley:2009bj}
A.~Buckley, H.~Hoeth, H.~Lacker, H.~Schulz and J.~E. von Seggern,
  \emph{{Systematic event generator tuning for the LHC}},
  \href{https://doi.org/10.1140/epjc/s10052-009-1196-7}{\emph{Eur. Phys. J.}
  {\bfseries C65} (2010) 331}
  [\href{https://arxiv.org/abs/0907.2973}{{\ttfamily 0907.2973}}].

\bibitem{Buckley:2010ar}
A.~Buckley, J.~Butterworth, L.~Lonnblad, D.~Grellscheid, H.~Hoeth, J.~Monk
  et~al., \emph{{Rivet user manual}},
  \href{https://doi.org/10.1016/j.cpc.2013.05.021}{\emph{Comput. Phys. Commun.}
  {\bfseries 184} (2013) 2803}
  [\href{https://arxiv.org/abs/1003.0694}{{\ttfamily 1003.0694}}].

\bibitem{Aad:2014xaa}
{\scshape ATLAS} collaboration, G.~Aad et~al., \emph{{Measurement of the
  $Z/\gamma^*$ boson transverse momentum distribution in $pp$ collisions at
  $\sqrt{s}$ = 7 TeV with the ATLAS detector}},
  \href{https://doi.org/10.1007/JHEP09(2014)145}{\emph{JHEP} {\bfseries 09}
  (2014) 145} [\href{https://arxiv.org/abs/1406.3660}{{\ttfamily 1406.3660}}].

\bibitem{Aad:2012wfa}
{\scshape ATLAS} collaboration, G.~Aad et~al., \emph{{Measurement of angular
  correlations in Drell-Yan lepton pairs to probe Z/gamma* boson transverse
  momentum at sqrt(s)=7 TeV with the ATLAS detector}},
  \href{https://doi.org/10.1016/j.physletb.2013.01.054}{\emph{Phys. Lett. B}
  {\bfseries 720} (2013) 32} [\href{https://arxiv.org/abs/1211.6899}{{\ttfamily
  1211.6899}}].

\bibitem{Aad:2011gj}
{\scshape ATLAS} collaboration, G.~Aad et~al., \emph{{Measurement of the
  transverse momentum distribution of Z/$\gamma*$ bosons in proton--proton
  collisions at $\sqrt{s}$=7 TeV with the ATLAS detector}},
  \href{https://doi.org/10.1016/j.physletb.2011.10.018}{\emph{Phys. Lett. B}
  {\bfseries 705} (2011) 415}
  [\href{https://arxiv.org/abs/1107.2381}{{\ttfamily 1107.2381}}].

\bibitem{Chatrchyan:2011wt}
{\scshape CMS} collaboration, S.~Chatrchyan et~al., \emph{{Measurement of the
  Rapidity and Transverse Momentum Distributions of $Z$ Bosons in $pp$
  Collisions at $\sqrt{s}=7$ TeV}},
  \href{https://doi.org/10.1103/PhysRevD.85.032002}{\emph{Phys. Rev. D}
  {\bfseries 85} (2012) 032002}
  [\href{https://arxiv.org/abs/1110.4973}{{\ttfamily 1110.4973}}].

\bibitem{Aad:2011fp}
{\scshape ATLAS} collaboration, G.~Aad et~al., \emph{{Measurement of the
  Transverse Momentum Distribution of $W$ Bosons in $pp$ Collisions at
  $\sqrt{s}=7$ TeV with the ATLAS Detector}},
  \href{https://doi.org/10.1103/PhysRevD.85.012005}{\emph{Phys. Rev. D}
  {\bfseries 85} (2012) 012005}
  [\href{https://arxiv.org/abs/1108.6308}{{\ttfamily 1108.6308}}].

\bibitem{Aad:2015auj}
{\scshape ATLAS} collaboration, G.~Aad et~al., \emph{{Measurement of the
  transverse momentum and $\phi ^*_{\eta }$ distributions of Drell--Yan lepton
  pairs in proton--proton collisions at $\sqrt{s}=8$ TeV with the ATLAS
  detector}}, \href{https://doi.org/10.1140/epjc/s10052-016-4070-4}{\emph{Eur.
  Phys. J. C} {\bfseries 76} (2016) 291}
  [\href{https://arxiv.org/abs/1512.02192}{{\ttfamily 1512.02192}}].

\bibitem{Sirunyan:2019bzr}
{\scshape CMS} collaboration, A.~M. Sirunyan et~al., \emph{{Measurements of
  differential Z boson production cross sections in proton-proton collisions at
  $ \sqrt{s} $ = 13 TeV}},
  \href{https://doi.org/10.1007/JHEP12(2019)061}{\emph{JHEP} {\bfseries 12}
  (2019) 061} [\href{https://arxiv.org/abs/1909.04133}{{\ttfamily
  1909.04133}}].

\bibitem{Gieseke:2007ad}
S.~Gieseke, M.~H. Seymour and A.~Siodmok, \emph{{A Model of non-perturbative
  gluon emission in an initial state parton shower}},
  \href{https://doi.org/10.1088/1126-6708/2008/06/001}{\emph{JHEP} {\bfseries
  06} (2008) 001} [\href{https://arxiv.org/abs/0712.1199}{{\ttfamily
  0712.1199}}].

\bibitem{Bahr:2008pv}
M.~Bahr et~al., \emph{{Herwig++ Physics and Manual}},
  \href{https://doi.org/10.1140/epjc/s10052-008-0798-9}{\emph{Eur. Phys. J.}
  {\bfseries C58} (2008) 639}
  [\href{https://arxiv.org/abs/0803.0883}{{\ttfamily 0803.0883}}].

\end{thebibliography}
\end{document}